\documentclass{ws-procs975x65}%
\usepackage{amsmath}
\usepackage{amsfonts}
\usepackage{amssymb}
\usepackage{graphicx}%
\def\beq{\begin{equation}}
\def\eeq{\end{equation}}
\begin{document}

\title{Gravity's Rainbow and Traversable Wormholes}
\author{Remo Garattini}

\address{Universit\`{a} degli Studi di Bergamo, \\
Dipartimento di Ingegneria e scienze applicate,\\
Viale Marconi,5 24044 Dalmine (Bergamo) ITALY\\
I.N.F.N. - sezione di Milano, Milan, Italy\\
$^*$E-mail:remo.garattini@unibg.it}

\author{Francisco S.~N.~Lobo}

\address{Instituto de Astrof\'isica e Ci\^encias do Espa\c{c}o, Universidade de Lisboa, \\
Faculdade de Ci\^encias, Campo Grande,\\ PT1749-016 Lisboa, Portugal.
\\
E-mail: fslobo@fc.ul.pt
}

\begin{abstract}
In the context of Gravity's Rainbow, we compute the graviton one-loop
contribution to a classical energy in a traversable wormhole background, by
considering the equation of state $p_{r} = \omega\rho$. The investigation is
evaluated by means of a variational approach with Gaussian trial wave
functionals. However, instead of using a regularization/renormalization
process, we use the distortion induced by Gravity's Rainbow to handle the divergences.

\end{abstract}

\bodymatter

\section{Introduction}

John A. Wheeler back in 1955 considered the possibility that spacetime could
be undergoing topological fluctuations at the Planck scale \cite{JAW}. This
changing spacetime is best known as the \textit{spacetime foam}, and can be a
model for the quantum gravitational vacuum. Indeed, Wheeler also considered
wormhole-type solutions, denoted as geons, obtained from the coupled equations
of electromagnetism and general relativity, as objects of the spacetime
quantum foam connecting different regions of spacetime at the Planck
scale~\cite{JAW}. In this proceedings, we consider the possibility of quantum
fluctuations in the context of Gravity's Rainbow \cite{GAC,GAC2,MagSmo}. The
latter consists of a distortion of the spacetime metric at energies of the
Planck energy, and for this purpose a general approach, denoted as deformed or
doubly special relativity, was developed in order to preserve the relativity
of inertial frames, maintain the Planck energy invariant and impose that in
the limit $E/E_{P}\rightarrow0$, the speed of a massless particle tends to a
universal and invariant constant, $c$.

More specifically, we explore the possibility that wormhole geometries are
sustained by their own quantum fluctuations, in the context of Gravity's
Rainbow \cite{RGFSNL}. We consider a semi-classical approach, where the
graviton one-loop contribution to a classical energy in a background spacetime
is computed through a variational approach with Gaussian trial wave
functionals. The energy density of the graviton one-loop contribution to a
classical energy in a wormhole background is considered as a self-consistent
source for wormholes \cite{Remo,Remo1,GL1}. In this semi-classical context, we
consider specific choices for the Rainbow's functions and find a plethora of
wormhole solutions, including non-asymptotically flat geometries and solutions
where the quantum corrections are exponentially suppressed, which provide
asymptotically flat wormhole geometries.

In fact, the possibility that quantum fluctuations induce a topology change,
in Gravity's Rainbow, has also been explored \cite{RGFL}. The energy density
of the graviton one-loop contribution, or equivalently the background
spacetime, was let to evolve, and consequently the classical energy was
determined. Note that the background metric was fixed to be Minkowskian in the
equation governing the quantum fluctuations, which behaves essentially as a
backreaction equation, and the quantum fluctuations were let to evolve. Then,
the classical energy, which depends on the evolved metric functions, was
evaluated. Analysing this procedure, a natural ultraviolet (UV) cutoff was
obtained, which forbids the presence of an interior spacetime region, and may
result in a multipy-connected spacetime. Thus, in the context of Gravity's
Rainbow, this process may be interpreted as a change in topology, and in
principle results in the presence of a Planckian wormhole.

\section{Gravity's Rainbow and the Equation of State}

\label{p1}

In Schwarzschild coordinates, the traversable wormhole metric can be cast into
the form \cite{MT}
\begin{equation}
ds^{2}=-\exp\left(  -2\phi\left(  r\right)  \right)  dt^{2}+ \left[
1-\frac{b\left(  r\right)  }{r}\right] ^{-1} dr^{2}+r^{2}d\Omega^{2}.
\label{metric}%
\end{equation}
where $\phi\left(  r\right)  $ is called the redshift function, while
$b\left(  r\right)  $ is called the shape function and where $d\Omega
^{2}=d\theta^{2}+\sin^{2}\theta d\phi^{2}$ is the line element of the unit
sphere. Using the Einstein field equation $G_{\mu\nu}=8\pi GT_{\mu\nu}$, we
obtain the following set of equations%
\begin{align}
\label{rhob}\rho\left(  r\right)   & = \frac{1}{8\pi G}\frac{b^{\prime}}{r^{2}%
},\\
p_{r}\left(  r\right)   & = \frac{1}{8\pi G}\left[  \frac{2}{r}\left(
1-\frac{b\left(  r\right)  }{r}\right)  \phi^{\prime}-\frac{b}{r^{3}}\right]
,\\
p_{t}\left(  r\right)   & = \frac{1}{8\pi G}\left(  1-\frac{b\left(  r\right)
}{r}\right)  \left[  \phi^{\prime\prime}+\phi^{\prime}\left(  \phi^{\prime
}+\frac{1}{r}\right)  \right]  -\frac{b^{\prime}r-b}{2r^{2}}\left(
\phi^{\prime}+\frac{1}{r}\right)  , \label{pt}%
\end{align}
where $\rho\left(  r\right)  $ is the energy density, $p_{r}\left(  r\right)
$ is the radial pressure, and $p_{t}\left(  r\right)  $ is the lateral
pressure. The conservation of the stress-energy tensor yields the following
relation
\begin{equation}
p_{r}^{\prime}=\frac{2}{r}\left(  p_{t}-p_{r}\right)  -\left(  \rho
+p_{r}\right)  \phi^{\prime}.
\end{equation}

When Gravity's Rainbow comes into play, the line element $\left(
\ref{metric}\right)  $ becomes\cite{MagSmo}
\begin{equation}
ds^{2}=-\exp\left(  -2\phi\left(  r\right)  \right)  \frac{dt^{2}}{g_{1}%
^{2}\left(  E/E_{P}\right)  }+\frac{dr^{2}}{\left(  1-b(r)/r\right)  g_{2}%
^{2}\left(  E/E_{P}\right)  }+\frac{r^{2}}{g_{2}^{2}\left(  E/E_{P}\right)
}d\Omega^{2}\, \label{dS}%
\end{equation}
and Einstein's Field Equations $\left(  \ref{rhob}\right) $--$\left(
\ref{pt}\right)  $ can be rearranged to give
\begin{align}
\label{b'g2}b^{\prime} & =\frac{8\pi G\rho\left(  r\right)  r^{2}}{g_{2}%
^{2}\left(  E/E_{P}\right)  },\\
\phi^{\prime} & =\frac{b+8\pi Gp_{r}r^{3}/g_{2}^{2}\left(  E/E_{P}\right)
}{2r^{2}\left(  1-b(r)/r\right)  }. \label{phig2}%
\end{align}

Now, we introduce the equation of state $p_{r}=\omega\rho$ \cite{Lobo:2005us},
and using Eq. $\left(  \ref{b'g2}\right)  $, then Eq. $\left(  \ref{phig2}%
\right)  $ becomes%
\begin{align}
\phi^{\prime} &  =\frac{b+8\pi G\left(  \omega g_{2}^{2}\left(  E/E_{P}%
\right)  b^{\prime}\left(  r\right)  /\left(  8\pi Gr^{2}\right)  \right)
r^{3}/g_{2}^{2}\left(  E/E_{P}\right)  }{2r^{2}\left(  1-b(r)/r\right)
}\nonumber\\
&  =\frac{b+\omega b^{\prime}r}{2r^{2}\left(  1-b(r)/r\right)  }.
\end{align}
Considering a constant redshift function yields the following condition%
\begin{equation}
b+\omega b^{\prime}r=0.\label{bb'}%
\end{equation}
which provides the solution%
\begin{equation}
b\left(  r\right)  =r_{0}\left(  \frac{r_{0}}{r}\right)  ^{\frac{1}{\omega}%
},\label{shape}%
\end{equation}
where we have used the condition $b\left(  r_{t}\right)  =r_{t}$. In this
situation, the line element $\left(  \ref{dS}\right)  $ becomes%
\begin{equation}
ds^{2}=-\frac{1}{g_{1}^{2}\left(  E/E_{P}\right)  }dt^{2}+\frac{dr^{2}%
}{1-\left(  \frac{r_{0}}{r}\right)  ^{1+\frac{1}{\omega}}g_{2}^{2}\left(
E/E_{P}\right)  }+\frac{r^{2}}{g_{2}^{2}\left(  E/E_{P}\right)  }d\Omega
^{2}.\label{line}%
\end{equation}
Note that the flaring-out condition entails the violation of the null energy
condition \cite{MT}, i.e., $p_{r}+\rho<0$, so that considering the equation of
state $p_{r}=\omega\rho$, the parameter is restricted by $\omega<-1$.

It is also possible to compute the proper radial distance modified by
Gravity's Rainbow%
\begin{align}
l\left(  r\right)   & = \pm\int_{r_{0}}^{r}\frac{dr^{\prime}}{\sqrt
{1-\frac{b_{\pm}\left(  r^{\prime}\right)  }{r^{\prime}}}}\nonumber\\
& =\pm\frac{r_{0}}{g_{2}\left(  E/E_{P}\right)  }\frac{2\omega}{\omega+1}%
\sqrt{\rho^{\left(  1+\frac{1}{\omega}\right)  }-1}\,_{2}F_{1}\ \left(
\frac{1}{2},\frac{1-\omega}{2\omega+2};\frac{3}{2};1-\rho^{\left(  1+\frac
{1}{\omega}\right)  }\right)  . \label{prd}%
\end{align}

It is interesting to note that Eq. $\left(  \ref{bb'}\right)  $ works also for
an inhomogeneous EoS. Indeed, the presence of the rainbow's function does not
affect the form of $\left(  \ref{bb'}\right)  $ except for an explicit
dependence on $r$ of the $\omega$ parameter, so that $b(r)+\omega(r)
b^{\prime}(r)r=0$ leads to the following general form of the shape function%
\begin{equation}
b(r)=r_{0} \, \exp\left[ - \int_{r_{0}}^{r} \frac{d\bar{r}}{\omega(\bar{r})
\bar{r}} \right]  . \label{form}%
\end{equation}

The situation appears completely different when a polytropic with an
inhomogeneous parameter $\omega$ is considered. Indeed, when the polytropic
EoS, i.e., $p_{r}=\omega\left(  r\right)  \rho^{\gamma}$, is plugged into Eq.
$\left(  \ref{phig2}\right)  $, one arrives%
\begin{align}
\phi^{\prime}  &  =\frac{b+8\pi Gp_{r}r^{3}/g_{2}^{2}\left(  E/E_{P}\right)
}{2r^{2}\left(  1-\frac{b\left(  r\right)  }{r}\right)  }=\frac{b+8\pi
G\left(  \omega\left(  r\right)  \rho^{\gamma}\right)  r^{3}/g_{2}^{2}\left(
E/E_{P}\right)  }{2r^{2}\left(  1-\frac{b\left(  r\right)  }{r}\right)
}\nonumber\\
&  =\frac{b+\left(  8\pi G\right)  ^{1-\gamma}\omega\left(  r\right)  \left(
b\left(  r\right)  ^{\prime}\right)  ^{\gamma}r^{3-2\gamma}g_{2}^{2\left(
\gamma-1\right)  }\left(  E/E_{P}\right)  }{2r^{2}\left(  1-\frac{b\left(
r\right)  }{r}\right)  }.
\end{align}
We can always impose that $\phi\left(  r\right)  =C$, but this means that%
\begin{equation}
b+\left(  8\pi G\right)  ^{1-\gamma}\omega\left(  r\right)  \left(  b\left(
r\right)  ^{\prime}\right)  ^{\gamma}r^{3-2\gamma}g_{2}^{2\left(
\gamma-1\right)  }\left(  E/E_{P}\right)  =0
\end{equation}
and a dependence on $g_{2}\left(  E/E_{P}\right)  $ appears. For this reason,
in this contribution, we will fix our attention on the case when $\omega$ is a constant.

\section{Self-sustained Traversable Wormholes, Gravity's Rainbow and Phantom
Energy}

\label{p2}

In this Section, we shall consider the formalism outlined in detail in Refs.
\cite{Remo,Remo1}, where the graviton one loop contribution to a classical
energy in a wormhole background is used. A traversable wormhole is said to be
\textquotedblleft\textit{self sustained}\textquotedblright\ if%
\begin{equation}
H_{\Sigma}^{(0)}=-E^{TT}, \label{SS}%
\end{equation}
where $E^{TT}$ is the total regularized graviton one loop energy and
$H_{\Sigma}^{(0)}$ is the classical term. When we deal with a spherically
symmetric line element, the classical Hamiltonian reduces to%
\begin{align}
H_{\Sigma}^{(0)} & =-\frac{1}{2G}\int_{r_{0}}^{\infty}\,\frac{dr\,r^{2}}%
{\sqrt{1-b(r)/r}}\,\frac{b^{\prime}(r)}{r^{2}g_{2}\left(  E/E_{P}\right)
}\nonumber\\
& =\frac{1}{2G}\int_{r_{0}}^{\infty}\,\frac{dr\,r^{2}}{\sqrt{1-b(r)/r}}%
\,\frac{b(r)}{r^{3}g_{2}\left(  E/E_{P}\right)  \omega},
\end{align}
where we have used the explicit expression of the scalar curvature in three
dimensions and the form of Eq. $\left(  \ref{bb'}\right)  $. Following Ref.
\cite{RGFSNL}, the self-sustained equation $\left(  \ref{SS}\right)  $
becomes
\begin{equation}
-\frac{b(r)}{2Gr^{3}g_{2}\left(  E/E_{P}\right)  \omega}=\frac{2}{3\pi^{2}%
}\left(  I_{1}+I_{2}\right)  \,, \label{ETT}%
\end{equation}
where the r.h.s. of Eq. $\left(  \ref{ETT}\right)  $ is represented by%
\begin{equation}
I_{1}=\int_{E^{\ast}}^{\infty}E\frac{g_{1}\left(  E/E_{P}\right)  }{g_{2}%
^{2}\left(  E/E_{P}\right)  }\frac{d}{dE}\left(  \frac{E^{2}}{g_{2}^{2}\left(
E/E_{P}\right)  }-m_{1}^{2}\left(  r\right)  \right)  ^{\frac{3}{2}}dE\,
\label{I1}%
\end{equation}
and%
\begin{equation}
I_{2}=\int_{E^{\ast}}^{\infty}E\frac{g_{1}\left(  E/E_{P}\right)  }{g_{2}%
^{2}\left(  E/E_{P}\right)  }\frac{d}{dE}\left(  \frac{E^{2}}{g_{2}^{2}\left(
E/E_{P}\right)  }-m_{2}^{2}\left(  r\right)  \right)  ^{\frac{3}{2}}dE\,,
\label{I2}%
\end{equation}
respectively. $E^{\ast}$ is the value which annihilates the argument of the
root while $m_{1}^{2}\left(  r\right)  $ and $m_{2}^{2}\left(  r\right)  $ are
two $r$-dependent effective masses. Of course, $I_{1}$ and $I_{2}$ are finite
for appropriate choices of the Rainbow's functions $g_{1}\left(
E/E_{P}\right)  $ and $g_{2}\left(  E/E_{P}\right)  $. With the help of the
EoS, one finds%
\begin{equation}
\left\{
\begin{array}
[c]{c}%
m_{1}^{2}\left(  r\right)  =\frac{6}{r^{2}}\left(  1-\frac{b\left(  r\right)
}{r}\right)  +\frac{3}{2r^{3}\omega}b\left(  r\right)  \left(  \omega+1\right)
\\
\\
m_{2}^{2}\left(  r\right)  =\frac{6}{r^{2}}\left(  1-\frac{b\left(  r\right)
}{r}\right)  +\frac{3}{2r^{3}\omega}b\left(  r\right)  \left(  \frac{1}%
{3}-\omega\right)
\end{array}
\right.
\end{equation}
and on the throat, $r=r_{0}$, the effective masses reduce to%
\begin{equation}%
\begin{tabular}
[c]{c}%
$m_{1}^{2}\left(  r_{0}\right)  =\frac{3}{2r_{0}^{2}\omega}\left(
\omega+1\right)  \qquad\left\{
\begin{array}
[c]{l}%
>0\qquad\mathrm{when\qquad}\omega>0\qquad\mathrm{or\qquad}\omega<-1\\
<0\qquad\mathrm{when\qquad}-1<\omega<0
\end{array}
\right.  $\\
\\
$m_{2}^{2}\left(  r_{0}\right)  =\frac{3}{2r_{0}^{2}\omega}\left(  \frac{1}%
{3}-\omega\right)  \qquad\left\{
\begin{array}
[c]{l}%
>0\qquad\mathrm{when\qquad}1/3>\omega>0\\
<0\qquad\mathrm{when\qquad}\omega>1/3\qquad\mathrm{or\qquad}\omega<0
\end{array}
\right.  $%
\end{tabular}
\ . \label{m1m2}%
\end{equation}

However, to have values of $\omega$ compatible with the flaring-out condition
and the violation of the null energy condition, only the case $\omega<-1$ is
allowed. It is easy to see that if we assume%
\begin{equation}
g_{1}\left(  E/E_{P}\right)  =1\qquad g_{2}\left(  E/E_{P}\right)  =\left\{
\begin{array}
[c]{c}%
1\qquad\mathrm{when}\qquad E<E_{P}\\
\\
E/E_{P}\qquad\mathrm{when}\qquad E>E_{P}\qquad
\end{array}
\right.  \,, \label{rel}%
\end{equation}
Eq. $\left(  \ref{ETT}\right)  $ becomes, close to the throat,%
\begin{equation}
-\frac{1}{2Gr_{0}^{2}\omega}=\frac{2}{\pi^{2}}\left(  \int_{\sqrt{m_{1}%
^{2}\left(  r\right)  }}^{E_{P}}E^{2}\sqrt{E^{2}-m_{1}^{2}\left(
r_{0}\right)  }dE\,+\int_{\sqrt{m_{2}^{2}\left(  r\right)  }}^{E_{P}}%
E^{2}\sqrt{E^{2}-m_{2}^{2}\left(  r_{0}\right)  }dE\right)  \,, \label{ETT1}%
\end{equation}
where $m_{1}^{2}\left(  r_{0}\right)  $ and $m_{2}^{2}\left(  r_{0}\right)  $
have been defined in Eq. $\left(  \ref{m1m2}\right)  $. Since the r.h.s. is
certainly positive, in order to have real solutions compatible with asymptotic
flatness, we need to impose $\omega<-1$, that it means that we are in the
Phantom regime. With this choice, the effective masses $\left(  \ref{m1m2}%
\right)  $ become, on the throat%
\begin{align}
m_{1}^{2}\left(  r_{0}\right)   & =\frac{3}{2r_{0}^{2}\omega}\left(
\omega+1\right)  \,,\\
m_{2}^{2}\left(  r_{0}\right)   & = -\frac{3}{2r_{0}^{2}\omega}\left(
\frac{1}{3}-\omega\right) \,,
\end{align}
and Eq. $\left(  \ref{ETT1}\right)  $ simplifies to%
\begin{align}
1=-\frac{4r_{0}^{2}\omega}{\pi^{2}E_{P}^{2}} \Bigg[ \int_{\sqrt{m_{1}^{2}
(r_{0}) }}^{E_{P}}E^{2}\sqrt{E^{2}-\frac{3}{2r_{0}^{2}\omega}\left(
\omega+1\right)  }dE\nonumber\\
+\int_{0}^{E_{P}}E^{2}\sqrt{E^{2}+\frac{3}{2r_{0}^{2}}\left\vert \frac
{1}{3\omega}-1\right\vert }dE \Bigg] \, \label{ETT2}%
\end{align}
The solution can be easily computed numerically and we find%
\begin{align}
-1 \geq\omega\geq-4.5,\nonumber\\
2.038 \geq x\geq1.083.\nonumber
\end{align}
Therefore we conclude that a wormhole which is traversable in principle, but
not in practice, can be produced joining Gravity's Rainbow and phantom energy.
Of course, the result is strongly dependent on the rainbow's functions which,
nonetheless must be chosen in such a way to give finite results for the one
loop integrals $\left(  \ref{I1}\right)  $ and $\left(  \ref{I2}\right)  $.

\section{Summary and further comments}

In this work, we have considered the possibility that wormhole geometries are
sustained by their own quantum fluctuations, but in the context of modified
dispersion relations. We considered different models regulated by the
Rainbow's functions to analyse the effect on the form of the shape function,
and found specific solutions for wormhole geometries. However, it is important
to point out that this approach presents a shortcoming mainly due to the
technical difficulties encountered. The variational approach considered in
this proceedings imposes a local analysis to the problem, namely, we have
restricted our attention to the behaviour of the metric function $b(r)$ at the
wormhole throat, $r_{t}$. Despite the fact that the behaviour is unknown far
from the throat, due to the high curvature effects at or near $r_{t}$, the
analysis carried out in this context should extend to the immediate
neighbourhood of the wormhole throat.

\section*{Acknowledgments}

FSNL was supported by the Funda\c{c}\~{a}o para a Ci\^{e}ncia e Tecnologia
(FCT) through the grants EXPL/FIS-AST/1608/2013, UID/FIS/04434/2013 and by a
FCT Research contract, with reference IF/00859/2012.

\end{document}